\documentstyle[12pt]{article}
\def\bra#1{\langle #1 |} \def\ket#1{| #1 \rangle}
\def\bracket#1#2{\bra{#1} #2 \rangle} 
\def\expect#1{\langle #1 \rangle} \def\d{^\dagger} \def\L{{\cal L}}

\def\Tr{{\rm Tr}}

\def\e{{\rm e}}
\def\half{\frac{1}{2}}

\begin{document}
\title{Quantum State Diffusion and Time Correlation Functions}

\author{Todd A. Brun \\ Physics Department, Queen Mary and Westfield College, \\
University of London, London~E1~4NS, England \\
and \\
Nicolas Gisin \\ Group of Applied Physics \\
University of Geneva, 1211 Geneva 4, Switzerland}

\date{\today}

\maketitle

\begin{abstract}
In computing the spectra of quantum mechanical systems one
encounters the Fourier transforms of time correlation functions, as
given by the quantum regression theorem for systems described by
master equations.  Quantum state diffusion (QSD) gives
a useful method of solving these problems by unraveling the master
equation into stochastic trajectories; but there is no generally
accepted definition of a time correlation function for a single
QSD trajectory.  In this paper we show how
QSD can be used to calculate these spectra directly; by formally
solving the equations which arise, we arrive at a natural definition
for a two-time correlation function in QSD, which depends explicitly
on both the stochastic noise of the particular trajectory and the
time of measurement, and which agrees in the mean with the ensemble
average definition of correlation functions.
\end{abstract}

\section{Introduction}

In quantum optics, a common experimental situation involves a system
in a cavity which is monitored by measuring the spectrum
of output photons.  In order to compare theory to experiment it is
necessary to calculate this output spectrum.

What is commonly found is that the intensity of the output spectrum is
given by the Fourier transform of two-time correlation functions of
system variables \cite{Mollow}.
In the case of a system isolated from its environment
(apart from the measurement process itself), these correlation functions
are simply the expectation values of
products of Heisenberg operators at different
times, e.g., $\expect{q(t_2) q(t_1)}$.

As it is difficult to completely isolate a system from the environment,
this approach does not always succeed.  In the case of a
quantum open system interacting continually with an external
reservoir, one instead describes the system by a master equation,
which - within the Markov approximation - can be written in Lindblad form:
\begin{equation}
{\dot \rho} = - i [H,\rho] + \sum_m \biggl( L_m \rho L_m\d
  - {1\over2} \{ L_m\d L_m, \rho \} \biggr).
\label{master_eqn}
\end{equation}
The $L_m$ are a set of environment operators which give the collective
effects of the environment, and we have taken $\hbar = 1$.

The density operator $\rho$ gives the probability for the
expected outcomes of measurements on the system.  In this case, one
can still calculate output spectra, and find Fourier
transforms of time correlation functions; these are no longer
the expectation values of products of Heisenberg operators in the system
Hilbert space alone, but are more
complicated objects whose form is given by the quantum regression theorem
\cite{QRT,Gardiner,Carmichael}.

Unfortunately, for
complicated systems equation (\ref{master_eqn}) can be very difficult to
solve either analytically or numerically.  In that case, it is often
advantageous to consider an unraveling of the master equation into
individual quantum trajectories, each one represented by a single state
at every moment in time.  One of these unraveling
techniques is quantum state diffusion (QSD) \cite{QSD}.
In this, the normalized vector
$\ket\psi$ representing the pure state of the system evolves
according to the QSD equation:
\[
d \ket\psi = - i H \ket\psi dt + \sum_m \biggl( \expect{L_m\d} L_m
  - {1\over2} L_m\d L_m - {1\over2} \expect{L_m\d} \expect{L_m} \biggr)
  \ket\psi dt
\]
\begin{equation}
+ \sum_m \biggl( L_m - \expect{L_m} \biggr) \ket\psi d\xi_m.
\label{qsd_eqn}
\end{equation}
This is an It\^o stochastic differential equation, in which the $d\xi_m$
represent independent complex Wiener processes.  These satisfy
\begin{equation}
M(d\xi_m) = M(d\xi_m d\xi_n) = 0, \ \ M(d\xi_m^* d\xi_n) = \delta_{mn} dt,
\end{equation}
where $M$ represents an ensemble average of the noise.  QSD reproduces
the master equation in the mean:
\begin{equation}
M(\ket\psi \bra\psi) = \rho.
\end{equation}
(This is what is meant by an unraveling of the master equation.)  Expectation
values for operators obey a similar relationship:
\begin{equation}
\expect{\hat O}_\rho = \Tr\{ {\hat O} \rho \} = M(\expect{\hat O}_\psi).
\label{qsd_expectation}
\end{equation}

The use of QSD as a practical algorithm to solve master equations has
been widely investigated \cite{QSD}.  This includes
calculations of output spectra in quantum optics \cite{Schack1}.  While it seems
reasonable that there should be a relationship between output spectra
and time correlation functions in QSD analogous to that in Schr\"odinger
or master equation dynamics, this has been difficult to show, as there
is no generally accepted definition of a time correlation function for a
single QSD trajectory.  Since QSD is framed in terms of a nonlinear
stochastic differential equation for an evolving state, it is not obvious
how to generalize this to products of operators at different times.

Gisin \cite{Gisin} has attempted to provide a Heisenberg picture for 
stochastically
evolving operators.  Sondermann \cite{Sondermann},
using the same stochastic equations,
has suggested another definition for
time correlation functions.  In both cases it is difficult to
relate this to the usual QSD formalism.  In this paper we attempt
to find a definition which arises from the QSD equation itself.

In section 2 we derive the output spectrum of a
quantum mechanical system, and show its relationship to time correlation
functions, as given by the quantum regression
theorem.

In section 3 we derive a quantum output
spectrum using QSD, and show
how it leads to a natural definition of a two-time correlation function
for a single QSD trajectory.  Unlike the correlation functions which arise
in the master equation derivation, these QSD functions depend explicitly
on the measurement time and the noise.  On average, however, these
dependencies vanish, and the result agrees with the usual definition of
a time correlation function.

In section 4 we look briefly at alternative unravelings, and discuss
the use of these techniques for the practical computation of spectra.
Finally, in section 5 we summarize our results and draw conclusions.

\section{Spectra and time correlation functions}

Consider a rather idealized model of an experimental set-up for
measuring the output spectrum of a quantum mechanical system.  Suppose we have
a system with Hamiltonian $H_1$ and Hilbert space ${\cal H}_1$
weakly coupled to an output
mode with Hamiltonian $H_2 = \omega b\d b$ and Hilbert space ${\cal H}_2$.
We assume weak coupling
to minimize the perturbation of the system evolution.  For
simplicity we assume a linear interaction:  $H_I = \epsilon q(b+b\d)$
where the operator $q$ represents a physical quantity of the system,
like position for example, $b$ and $b\d$ are the
annihilation and creation operators for the output mode, and $\epsilon$ is
small.

The total Hamiltonian of the system plus output mode is
\begin{equation}
H = H_1 + H_2 + H_I
\label{hamiltonian}
\end{equation}
and operates on the combined Hilbert space ${\cal H}_1 \otimes {\cal H}_2$.
Let the system degrees of freedom also be coupled to an
environment, described by a set of environment operators $L_m$ acting on
${\cal H}_1$.
The system plus output mode
obeys the master equation (\ref{master_eqn}) with Hamiltonian
(\ref{hamiltonian}).

If the output mode is initially in the ground state, so that the initial
density matrix is of the form $\rho = \rho_{00}\otimes\ket0 \bra0$,
we can approximate at later times
\begin{equation}
\rho \approx \rho_{00} \otimes \ket0 \bra0 + \rho_{01} \otimes \ket0 \bra1
  + \rho_{10} \otimes \ket1 \bra0 + \rho_{11} \otimes \ket1 \bra1
  + O(\epsilon^3),
\end{equation}
where the $\rho_{ij}$ are time-dependent operators on ${\cal H}_1$;
we have explicitly separated the system and output mode degrees of
freedom and neglected all but the first excited state of the output mode.
We then have $\rho_{01}, \rho_{10} \sim \epsilon$ and
$\rho_{11} \sim \epsilon^2$, and can rewrite equation (\ref{master_eqn})
in the form
\begin{eqnarray}
{\dot \rho_{00}} && = \L \rho_{00} + O(\epsilon^2), \nonumber\\
{\dot \rho_{01}} && = \L \rho_{01} + i \epsilon \rho_{00} q
  + i \omega \rho_{01} + O(\epsilon^3), \nonumber\\
{\dot \rho_{10}} && = \L \rho_{10} - i \epsilon q \rho_{00}
  - i \omega \rho_{10} + O(\epsilon^3), \nonumber\\
{\dot \rho_{11}} && = \L \rho_{11} - i \epsilon q \rho_{01}
  + i \epsilon \rho_{10} q + O(\epsilon^4).
\label{component_eqns}
\end{eqnarray}
$\L$ is the time evolution superoperator
restricted to the system degrees of freedom:
\begin{equation}
\L \rho_{ij} = - i [H_1, \rho_{ij}] + \sum_m \biggl( L_m \rho_{ij} L_m\d
  - \half \{ L_m\d L_m, \rho_{ij} \} \biggr).
\label{Lsystem}
\end{equation}
To lowest order, $\rho_{00}$ evolves according to a normal master
equation, with no reference to the output mode at all.  The other components
$\rho_{ij}$ represent the weak signal transmitted
via the interaction with the output mode.

We can solve these equations:
\begin{equation}
\rho_{00}(t) = S_0^t \rho_{00}(0);
\end{equation}
\begin{equation}
\rho_{01}(t) = i \epsilon \int_0^t S^t_{t'}
  (\rho_{00}(t') q ) \e^{i\omega(t-t')} dt'
  = \rho_{10}\d(t)
\end{equation}
\begin{equation}
\rho_{11}(t) = -i \epsilon \int_0^t S^t_{t'}
  (q \rho_{01}(t') ) dt' + {\rm h.c.}
\label{output_mode}
\end{equation}
\[
 = \epsilon^2 \int_0^t \int_0^{t'} S^t_{t'} (q S^{t'}_{t''}
  ( ( S_0^{t''} \rho_{00}(0) ) q ) ) \e^{i \omega (t' - t'')} dt'' dt'
  + {\rm h.c.},
\]
where $S_{t_1}^{t_2} = \exp(\L(t_2 - t_1) )$ is the time evolution
superoperator from time $t_1$ to time $t_2$ given by (\ref{Lsystem}).
The excitation of the output mode is what is measured, so we are interested
in the expectation value $\expect{b\d b} = \Tr \{ b\d b \rho \}
= \Tr \{ \rho_{11} \}$.  We find this by taking the trace of
equation (\ref{output_mode}).  Equation (\ref{master_eqn}) preserves
the trace, so in taking the trace of (\ref{output_mode}) the factor
$\exp(\L(t-t'))$ has no effect, and may be dropped.  Thus,
\begin{equation}
\expect{b\d b}(t)
  = \epsilon^2 \int_0^t \int_0^{t'} \Tr \biggl\{ q S^{t'}_{t''}
  ( (S_0^{t''} \rho_{00}(0) ) q) \biggr\}  \e^{i \omega (t' - t'')} dt'' dt'
  + {\rm c.c.}
\label{spectrum}
\end{equation}
\[
  = \epsilon^2 \int_0^t \int_0^{t'} \Tr \biggl\{ q S^{t'}_{t''}
  ( \rho_{00}(t'') q ) \biggr\}  \e^{i \omega (t' - t'')} dt'' dt'
  + {\rm c.c.}
\]
\[
  = \epsilon^2 \int_0^t \int_0^{t'}
  \e^{i \omega (t' - t'')}
  \expect{q(t')q(t'')}_{\rm QRT} dt'' dt'
  + {\rm c.c.},
\]
where the last equality defines the two-time correlation
function $\expect{q(t')q(t'')}_{\rm QRT}$. Note that
$\expect{q(t')q(t'')}_{\rm QRT}$ is determined by the evolution
operator $S_0^t$ of the reduced mixed state $\rho_{00}$,
in accordance with the quantum regression theorem \cite{QRT}.  We see that
the expected output has the form of a Fourier transform of a
two-time correlation function; this is just like the result in classical
physics \cite{Classical}.
Examining output modes at different frequencies $\omega$
gives the spectrum of the system.


\section{Spectra and time correlations in QSD}

We can calculate the results of section 2 with QSD by using relation
(\ref{qsd_expectation}).  We solve the QSD equation
(\ref{qsd_eqn}) for a state in the combined Hilbert space
${\cal H} = {\cal H}_1 \otimes {\cal H}_2$, starting with an initial
condition $\ket\Psi = \ket{\psi_0} \ket0$.
By averaging $\expect{b\d b}_{\Psi(t)}$
over many trajectories we reproduce (\ref{spectrum}).
This was done in \cite{Schack1} for the case of second harmonic generation.

How can we interpret $\expect{b\d b}_{\Psi(t)}$
for a {\it single} trajectory?  Let's examine a little more closely the
evolution (\ref{qsd_eqn}).  We can separate the components of $\Psi$
\begin{equation}
\ket{\Psi(t)} = \ket{\phi_0(t)}\ket0 + \ket{\phi_1(t)}\ket1 + O(\epsilon^2),
\end{equation}
where we again neglect all excited states of the output mode above
the first.  The initial condition is 
\begin{equation}
\ket{\phi_0(0)} = \ket{\psi_0},\ \  \ket{\phi_1(0)} = 0.
\end{equation}
The QSD equation becomes a pair of coupled equations
\begin{eqnarray}
d \ket{\phi_0} && = - i H_1 \ket{\phi_0} dt
  + \sum_m \biggl( \expect{L_m\d}_{\phi_0} L_m
  - {1\over2} L_m\d L_m
  - {1\over2} \expect{L_m\d}_{\phi_0} \expect{L_m}_{\phi_0}
  \biggr) \ket{\phi_0} dt \nonumber\\
&& + \sum_m \biggl( L_m - \expect{L_m}_{\phi_0} \biggr) \ket{\phi_0} d\xi_m
  + O(\epsilon^2),
\label{system_eqn}
\end{eqnarray}
\begin{eqnarray}
d\ket{\phi_1} && = - i H_1 \ket{\phi_1} dt
  + \sum_m \biggl( \expect{L_m\d}_{\phi_0} L_m
  - {1\over2} L_m\d L_m
  - {1\over2} \expect{L_m\d}_{\phi_0} \expect{L_m}_{\phi_0}
  \biggr) \ket{\phi_1} dt \nonumber\\
&& + \sum_m \biggl( L_m - \expect{L_m}_{\phi_0} \biggr) \ket{\phi_1} d\xi_m
  - i \omega \ket{\phi_1} dt
  - i \epsilon q \ket{\phi_0} dt + O(\epsilon^3),
\label{output_eqn}
\end{eqnarray}
Note that (\ref{system_eqn}) is identical to the usual QSD
equation for the system alone
to first order in $\epsilon$.  Just as in section 2, this approximation
gains an extra order of $\epsilon$ in accuracy for free.
Thus, the system degrees
of freedom have the usual quantum state diffusion behavior, essentially
uninfluenced by the interaction with the output mode.
Exactly as in (\ref{component_eqns}), we have an almost unperturbed
system, with a weak signal transmitted to the outside world.

Equation (\ref{output_eqn}) is interesting, in that
all of the expectation values in this equation
are calculated with respect to $\ket{\phi_0}$, so this is a driven
linear equation with time-dependent coefficients.  Given
the solution $\ket{\phi_0(t)}$,
we can find $\ket{\phi_1(t)}$, at least in principle.

We formally integrate (\ref{system_eqn}) to get
\begin{equation}
\ket{\phi_0(t_2)} = T(\xi,\psi_0)_{t_1}^{t_2} \ket{\phi_0(t_1)},
\end{equation}
where $T(\xi,\psi_0)_{t_1}^{t_2}$ is the time-evolution operator from
time $t_1$ to time $t_2$.  It explicitly depends on the noise $\xi$ and
the initial state $\ket{\psi_0}$, since the QSD equation is nonlinear.
Given this time-evolution operator, the solution
to (\ref{output_eqn}) is
\begin{equation}
\ket{\phi_1(t)} =
  - i \epsilon \int_0^t T(\xi,\psi_0)_{t'}^t q \ket{\phi_0(t')}
  \e^{-i \omega (t - t')} dt'.
\end{equation}
The output spectrum is then
\begin{equation}
\expect{b\d b}_\Psi(t) =
  \epsilon^2 \int_0^t \int_0^t \bra{\psi_0} {T\d}_0^{t'} q
  {T\d}_{t'}^t T_{t''}^t q T_0^{t''} \ket{\psi_0}
  \e^{i\omega(t'-t'')} dt' dt'',
\label{qsd_spectrum}
\end{equation}
where the $\xi$ and
$\psi_0$ have been suppressed for conciseness.  This expression resembles
a Fourier transform of some kind of correlation function, just as was
the case in section 2; but this correlation function is defined for a
single QSD trajectory.

We can bring this rather closer to the treatment in section 2 by introducing
the projector
\begin{eqnarray}
P_\Psi && = \ket\Psi \bra\Psi, \nonumber\\
&& = P_{00} \ket0\bra0 + P_{01} \ket0\bra1 + P_{10} \ket1\bra0
  + P_{11} \ket1\bra1,
\end{eqnarray}
where these {\it partial projectors} $P_{ij}$ are operators on ${\cal H}_1$.
These are related to our earlier treatment by
\begin{equation}
P_{00} = \ket{\phi_0}\bra{\phi_0},\ \ P_{01} = \ket{\phi_0}\bra{\phi_1},\ \
P_{10} = \ket{\phi_1}\bra{\phi_0},\ \ P_{11} = \ket{\phi_1}\bra{\phi_1},
\end{equation}
and to section 2 by
\begin{equation}
\rho_{ij} = M(P_{ij}).
\end{equation}
These $P_{ij}$ are not themselves projectors; however, $P_{00}^2 =
P_{00} + O(\epsilon^2)$, and therefore can be considered a
projector to good approximation.

We define a time evolution superoperator for these partial projectors
\begin{equation}
S(\xi,\psi_0)_{t_1}^{t_2} P = T(\xi,\psi_0)_{t_1}^{t_2}~P~
  {T\d}(\xi,\psi_0)_{t_1}^{t_2}.
\label{qsd_superoperator}
\end{equation}
This is related to the superoperator for the master equation by
\begin{equation}
M(S(\xi,\psi_0)_{t_1}^{t_2} )~P_{00} = S_{t_1}^{t_2}~P_{00}
 = \exp( \L(t_2 - t_1) )~P_{00}.
\end{equation}
We shall need the following solutions:
\begin{equation}
P_{00}(t) = S(\xi,\psi_0)_0^t P_{00}(0),
\end{equation}
\begin{equation}
P_{01}(t) = i\epsilon \int_0^t S(\xi,\psi_0)_{t'}^t (P_{00}(t') q)
  \e^{i\omega(t-t')} dt' = P_{10}\d(t)
\end{equation}
%
%
%

In order to compute the output signal $\expect{b\d b}_{\psi(t)}$
we make use of the equality:
\begin{equation}
P_{11}(t) = P_{10} P_{01}.
\end{equation}
In this way, we obtain the following expression for the output signal:
\begin{equation}
\expect{b\d b}_{\psi(t)} = Tr\{P_{10} P_{01}\}
\end{equation}
\[
= \epsilon^2 \int_0^t dt'
  \Tr \biggl\{ S(\xi,\psi_0)_{t'}^t (P_{00}(t') q)\d \int_0^t dt''
  S(\xi,\psi_0)_{t''}^t (P_{00}(t'') q) \biggr\} \e^{i\omega(t' - t'')},
\]
\[
= \epsilon^2 \int_0^t dt'
  \Tr\biggl\{ T_{t'}^t q P_{00}(t') {T_{t'}^t}\d \int_0^t dt''
  T_{t''}^t P_{00}(t'') q {T_{t''}^t}\d \biggr\} \e^{i\omega(t' - t'')}.
\]
Note that the two central operators ${T_{t'}^t}\d T_{t''}^t$ can not be
replaced with ${T_{t'}^{t''}}\d$.

Accordingly, in analogy with (\ref{spectrum}) we define a
{\it measurement-dependent} two-time correlation function for QSD:
\begin{equation}
C({\hat O}_2,t_2; {\hat O}_1,t_1; t)
\equiv  \Tr \biggl\{ S(\xi,\psi_0)_{t_2}^t(\hat O_2 P_{00}(t_2) )~
  S(\xi,\psi_0)_{t_1}^t( P_{00}(t_1)\hat O_1 ) \biggr\}
\label{qsd_correlation}
\end{equation}
Note that this assumes
$t_1 < t_2$; for $t_1 > t_2$ an analogous expression can be formed.
This function $C$ in QSD has the nice feature that it is the trace of
a product of two operators, each involving one of the times $t_1$ and $t_2$,
similar to correlation functions of
classical stochastic processes (which are products of the random variables at
different times \cite{Classical}). However, it also has the strange feature 
that the 
``final''
time $t$ appears in this definition.  This is why we term $C$ a
measurement-dependent correlation function, rather than a true
correlation function.
Note that in perfect analogy with the quantum regression
theorem, $C({\hat O}_2,t_2; {\hat O}_1,t_1; t)$ is
determined by the evolution operator $S(\xi,\psi_0)_{t_1}^{t_2}$
of the reduced system's pure state $\ket{\phi_0(t)}$.

It remains to establish the relation between the function
$C({\hat O}_2,t_2; {\hat O}_1,t_1; t)$ and the QRT
correlation function introduced in section 2. For this purpose
we note that $P_{11} = \ket{\phi_1}\bra{\phi_1}$ provides an equivalent
expression for the output signal:
\[
\expect{b\d b}_\Psi(t) = \Tr\{P_{11}\}
\]
\begin{equation}
 = \epsilon^2 \int_0^t \int_0^{t'} \Tr \biggl\{ S(\xi,\psi_0)_{t'}^t
  \biggl( q S(\xi,\psi_0)^{t'}_{t''}
  ( P_{00}(t'') q ) \biggl) \biggr\}  \e^{i \omega (t' - t'')} dt' dt''
  + {\rm c.c.}
\label{TimeCorr2}
\end{equation}
Using this relation (\ref{TimeCorr2}) we can derive an equivalent form of 
the function $C$:
\[
C({\hat O}_2,t_2; {\hat O}_1,t_1; t) =
\]
\begin{equation}
\Tr \biggl\{ S(\xi,\psi_0)_{t_2}^t \biggl( {\hat O}_2 S(\xi,\psi_0)_{t_1}^{t_2}
  ( P_{00}(t_1) {\hat O}_1 ) \biggl) \biggr\},
\label{qsd_correlation2}
\end{equation}
Note that in section 2 it was possible to
remove the time evolution superoperator $S_{t'}^t$, as it did not affect the
trace.  This is not true in the single trajectory case,
as explained above; so the function $C$ depends explicitly
on time, as mentioned above.
Nevertheless, taking the mean over the noise of
$C({\hat O}_2,t_2; {\hat O}_1,t_1; t)$,
one recovers the
correlation function derived for mixed states from the quantum regression
theorem:
\begin{equation}
M(C({\hat O}_2,t_2; {\hat O}_1,t_1; t)) =
  \expect{{\hat O}_2(t_2) {\hat O}_1(t_1) }_{\rm QRT}.
\label{theorem}
\end{equation}
Note that the dependence on the final time $t$ vanishes in the mean.

While the definition (\ref{qsd_correlation}) arises naturally in this
derivation, and has the correct average behavior, its dependence on
$t$ remains a puzzling and rather annoying feature.  It is possible to
make a different definition,
closely related to that of (\ref{qsd_correlation}),
which avoids this problem.
Since the operator $S(\xi,\psi_0)_{t_2}^t$ in the definition
(\ref{qsd_correlation2}) vanishes in the mean,
we can define a true two-time correlation function:
\begin{equation}
\expect{{\hat O}_2(t_2) {\hat O}_1 (t_1)}_{\rm QSD} \equiv
\Tr \biggl\{ {\hat O}_2 S(\xi,\psi_0)_{t_1}^{t_2}
  ( P_{00}(t_1) {\hat O}_1 ) \biggr\}.
\label{qsd_correlation3}
\end{equation}
In this way the ``final time'' $t$ disappears and the correlation
function is formally
identical to the QRT case (but for vectors instead of matrices).
Moreover, the correlation function then appears as a scalar product of
two vectors:
\begin{equation}
\expect{{\hat O}_2(t_2) {\hat O}_1 (t_1)}_{\rm QSD} =
\bracket{T_{t_1}^{t_2}\hat O_1 \phi_0(t_1)}{\hat O_2 \phi_0(t_2)}.
\end{equation}

One can consider this new definition to be an average over ``future'' noise,
i.e., noise after $t_2$.
\begin{equation}
\expect{{\hat O}_2(t_2) {\hat O}_1(t_1)}_{\rm QSD} =
  M(C({\hat O}_2,t_2; {\hat O}_1,t_1; t))_{\xi(t'), t_1 < t_2 < t' < t}.
\end{equation}
Certainly, this once again reproduces the
QRT correlation function in the mean, and gives exactly the same
$\expect{b\d b}_{\psi(t)}$  for a single trajectory:
\begin{equation}
M(\expect{{\hat O}_2(t_2) {\hat O}_1(t_1)}_{\rm QSD}) =
  \expect{{\hat O}_2(t_2) {\hat O}_1(t_1) }_{\rm QRT}.
\label{theorem2}
\end{equation}

While this definition removes the dependence on the final time $t$, this
two-time correlation function still has some interesting features, related
to the fact that the QSD time-evolution superoperator (\ref{qsd_superoperator})
does not preserve the trace.  In particular, we note that if the second
operator ${\hat O}_2$ is the identity ${\hat I}$, then
\begin{equation}
\expect{{\hat I}(t_2) {\hat O}_1(t_1)}_{\rm QSD} \ne \expect{{\hat O}_1(t_1)},
\end{equation}
contrary to the case of the QRT.  If ${\hat O}_1 = {\hat I}$ no such
difficulty arises.  Since unravelings of the master equation do not
generically preserve the trace, this feature will arise in any similar
derivation.  This might be one argument for using an unraveling specifically
chosen to preserve the trace, such as that used by Gisin in his discussion
of a Heisenberg picture for QSD \cite{Gisin,Sondermann}.   In any case,
the desired relation does hold in the mean:
\begin{equation}
M(\expect{{\hat I}(t_2) {\hat O}_1(t_1)}_{\rm QSD})
  = M(\expect{{\hat O}_1(t_1)}).
\end{equation}
Fortunately, correlation functions of the above form would not arise in
any physically reasonable measurement scheme, since ${\hat I}$ does not
describe an interaction.

The definition of the two-time correlation function (\ref{qsd_correlation3}),
and its relation
to the quantum regression theorem (\ref{theorem2}) are
the main results of this article. The central line of arguments
can be summarized as follows:
\begin{equation}
\rho_{11}=M(P_{11})=M(P_{10}P_{01})\neq 
M(P_{10})M(P_{01})=\rho_{10}\rho_{01}
\end{equation}

\section{Alternative unravelings and practical calculations}

While the definition (\ref{qsd_correlation3})
arises naturally from the QSD analysis of this problem, it is not the
only possible definition of a time correlation function for a single
trajectory; indeed, there are an infinite number of such definitions,
corresponding to different unravelings \cite{Diosi}.

One such alternative unraveling has already been proposed by
Gisin as a possible definition of a time-correlation function for
individual trajectories, as well as providing a sort of Heisenberg
picture corresponding to QSD \cite{Gisin}.
Here we will suggest yet another such
alternative unraveling.

In particular, rather than solving the Lindblad master equation itself
and then solving for the output spectrum, one might instead begin
with the expression (\ref{spectrum}) and attempt to unravel the
time-correlation function directly.

What form would such an unraveling take?  In QSD, one unravels the
density operator evolution into many trajectories, each consisting
of a single state, and with a mean $\rho = M(\ket\psi \bra\psi)$.
This works because $\rho$ is hermitian:  $\rho = \rho\d$.
For a time correlation function, this is no longer sufficient.
The form given by the quantum regression theorem is
\begin{equation}
  \expect{q(t_2)q(t_1)}_{\rm QRT}
= \Tr \biggl\{ q S^{t_2}_{t_1}
  ( \rho(t_1) q ) \biggr\},
\label{qrt_correlation}
\end{equation}
where $S^{t_2}_{t_1}$ is the time-evolution superoperator defined by
the master equation (\ref{master_eqn}); however, the ``initial state''
$\rho(t_1) q$ is {\it not} hermitian.  Therefore
the evolution from $t_1$ to $t_2$ cannot be unraveled
in terms of pure states.

One can, however, consider a pair of vectors
$\ket\psi$ and $\ket\phi$ such that $M(\ket\psi \bra\phi)$ {\it does}
reproduce the correct evolution.  Such a {\it diad} equation is quite
analogous to the usual QSD equation.

One pair of coupled equations that do the job are
\begin{eqnarray}
\ket{d\psi} = \biggl( - i H + \expect{L\d}_\phi L
  - {1\over2} L\d L
  - {1\over2} \expect{L\d}_\phi \expect{L}_\psi \biggr) \ket\psi dt
  + (L - \expect{L}_\psi ) \ket\psi d\xi, \nonumber\\
\bra{d\phi} =  \bra\phi \biggl( i H + L\d \expect{L}_\psi
  - {1\over2} L\d L
  - {1\over2} \expect{L\d}_\phi \expect{L}_\psi \biggr) dt
  + \bra\phi (L\d - \expect{L\d}_\phi ) d\xi^*.
\label{diad}
\end{eqnarray}
It is not hard to show that
\begin{equation}
M(d \ket\psi \bra\phi) = \L (\ket\psi\bra\phi),
\end{equation}
so this has the correct evolution in the mean.
If $\ket\psi = \ket\phi$ then these coupled equations
reduce to the ordinary QSD equation (\ref{qsd_eqn}).  Note that
in general, however, the normalization of $\ket\psi$ and $\ket\phi$
is {\it not} preserved.

The technique for calculating time correlation functions is as follows.
If the initial density matrix is a pure state,
$\rho(0) = \ket{\psi_0} \bra{\psi_0}$,
then one begins with both states equal $\ket\psi = \ket\phi = \ket{\psi_0}$,
and evolves them to time $t_1$ according to
(\ref{diad}), which is equivalent to the QSD equation
(\ref{qsd_eqn}).  At time $t_1$, multiply
$\bra{\phi(t_1)} \rightarrow \bra{\phi(t_1)} q$
and continue to evolve the diad
according to (\ref{diad}).  At time $t_2$, multiply
$\ket{\psi(t_2)} \rightarrow q \ket{\psi(t_2)}$ and take the trace.
The mean over many such trajectories equals the time-correlation
function (\ref{qrt_correlation}).

This pair of equations (\ref{diad}) shares many properties in common
with the QSD equation (\ref{qsd_eqn}), but unlike QSD is not uniquely
defined.  Since only the composite diad $\ket\psi \bra\phi$ is important,
the norm and phase can be shifted arbitrarily between these two states.
Di\'osi has likened this to a gauge freedom \cite{Diosi}.
Many such pairs of equations
are therefore possible, as well as others with properties radically
different from QSD; just as QSD is one of many unravelings
of the master equation, albeit with unique symmetry properties.
To solve for output spectra, of course, one must still
Fourier transform the calculated time correlation functions.

In fact, one can see that there are two distinct approaches
to computing output spectra using QSD.  One is to use the definition
(\ref{qsd_correlation3}) or alternative definitions such as
(\ref{diad}) to calculate the time-correlation function, averaging
over many runs, and taking the Fourier transform
(\ref{spectrum}).

Alternatively, one can solve the QSD equation for the entire system
plus output mode; the state is then in the larger Hilbert space
${\cal H}_1 \otimes {\cal H}_2$.  This is essentially the approach
taken by Schack et al. \cite{Schack1}, who have also shown that
these techniques can be used to calculate other quantities of
interest, such as the spectrum of squeezing.

Both approaches appear to have their advantages and disadvantages, and
to be roughly equal in computational difficulty.  It is likely that
the best approach will vary from problem to problem.

Note also that if we had unraveled the master equation (\ref{master_eqn})
using an unraveling other than QSD, an exactly analogous argument would
have followed.  We would be able to resolve the equation for
the system plus output mode into a pair of coupled equations, one
corresponding to the unperturbed evolution of the system alone, the
other to an output signal completely driven by the system.  Thus,
this type of argument could be used to define a notion of time correlation
functions for single trajectories in any unraveling, such as the Quantum
Monte Carlo techniques or the orthojumps of Di\'osi
\cite{Carmichael,Dalibard,Diosi2}.

These Quantum Monte Carlo (or Quantum Jump) techniques deserve further
comment, as there have already been a number of papers published
on their use in the calculation of time correlation functions
\cite{Mollmer,Gardiner,Dum}.  These are relevant to our current discussion,
since it can be shown that the equations for Quantum Jumps become
identical to those for QSD in the case of heterodyne measurements
\cite{Heterodyne}.

The treatment of Gardiner, Parkins, and Zoller \cite{Gardiner} is particularly
interesting in this context.  Their definition of the time correlation
function involves defining an auxiliary vector $\ket{\beta,t}$ which
is driven by the evolution of the quantum jump vector $\ket{\phi,t}$,
but does not in turn affect it.

As stated in that paper \cite{Gardiner} their equations are rather
different.  In particular, they are considering only real noise, and
have not gone to the weak coupling limit we have assumed in this paper,
where the output has a negligibly weak effect on the system over short
times.  Wiseman and Milburn \cite{Heterodyne} have generalized this treatment
to consider the case of heterodyne measurements, introducing complex noise
and showing that this limit is exactly equivalent to the Quantum State
Diffusion equation.  In the limit of heterodyne measurement with weak
coupling to the external mode,
the equation for $\ket{\phi,t}$ becomes the QSD equation (\ref{qsd_eqn}),
and the two vectors $\ket{\phi,t}$ and $\ket{\beta,t}$ obey a pair
of coupled equations identical to (\ref{system_eqn}) and
(\ref{output_eqn}).  The output spectrum is given by the mean of
$\bracket{\beta,t}{\beta,t}$, just as in (\ref{qsd_spectrum}).  From
this, one could follow an argument exactly analogous to that of
section 3 in this paper to arrive at a definition of a two-time correlation
function identical to (\ref{qsd_correlation3}).

\section{Conclusions}
In calculating quantum optical spectra, a common approach is to calculate
the quantum time correlation function and derive the spectrum by taking
its Fourier transform.  This time correlation function has a form given
by the quantum regression theorem, and requires a solution of the master
equation.

Quantum state diffusion provides in many cases an efficient method of
solving the master equation.  But hitherto, an appropriate definition of
the time correlation function for a single QSD trajectory has been lacking,
making QSD less useful for the calculation of spectra.

In this paper we have derived such a definition in a straightforward way,
quite analogous to the derivation in the
case of the full master equation.  This correlation function in the mean
has the form given by the quantum regression theorem, and the equations
for it are very close to the original QSD equation, up to $O(\epsilon^2)$
in the interaction strength.

This definition can be used
as a practical numerical tool in computing quantum optical spectra.
Other possible definitions and their potential
for practical use have been briefly discussed.

Finally, let us discuss the meaning of our result in
the simple case of a damped harmonic
oscillator at zero temperature. The stationary solution is the ground state.
Hence, once the
system has reached this state, nothing happens, and it is clear that the
spectrum of the damped oscillator is not contained in the evolution of its
state vector.
Nevertheless, the quantum regression theorem tells us that
all spectra, in particular the one corresponding to the position fluctuation,
are contained in the evolution operator for the corresponding master equation.
Similarly, the results presented here tell us that the spectra are also
contained in the stochastic evolution operator $T(\xi,\psi_0)_{t_1}^{t_2}$
of the QSD description of the damped oscillator. The physics behind this is
that whenever a spectrum is measured, the system's environment is changed,
hence its dynamics is perturbed.
For example, to measure the spectrum of position fluctuations,
something like weak position measurements have to be applied,
and the ground state is no longer stationary \cite{Salama}.
However, in contrast to standard quantum
measurements, this perturbation can be made arbitrarily small (corresponding
to small amplitudes of the measured spectrum) over an arbitrarily
long period of time.
To first order, the system's
evolution is unaffected, but its states acts like a source for the signal,
as reflected by our equations (\ref{system_eqn}) and (\ref{output_eqn}).
Hence the unaffected evolution operator (Liouville operator or QSD
propagator or other stochastic propagator \cite{Carmichael,Mollmer})
contains the information about spectra that cannot actually be measured
without affecting (i.e., weakly perturbing) the system's evolution.

\section*{Acknowledgments}
We would like to acknowledge many useful conversations with Lajos Di\'osi,
Ian Percival, Marco Rigo, R\"udiger Schack, Walter Strunz and Tim Spiller.
We would also like to thank our referee, who provided a great deal of
important feedback and helped us to clarify our results.  This work was
supported in part by the Swiss National Science Foundation.


\vfil

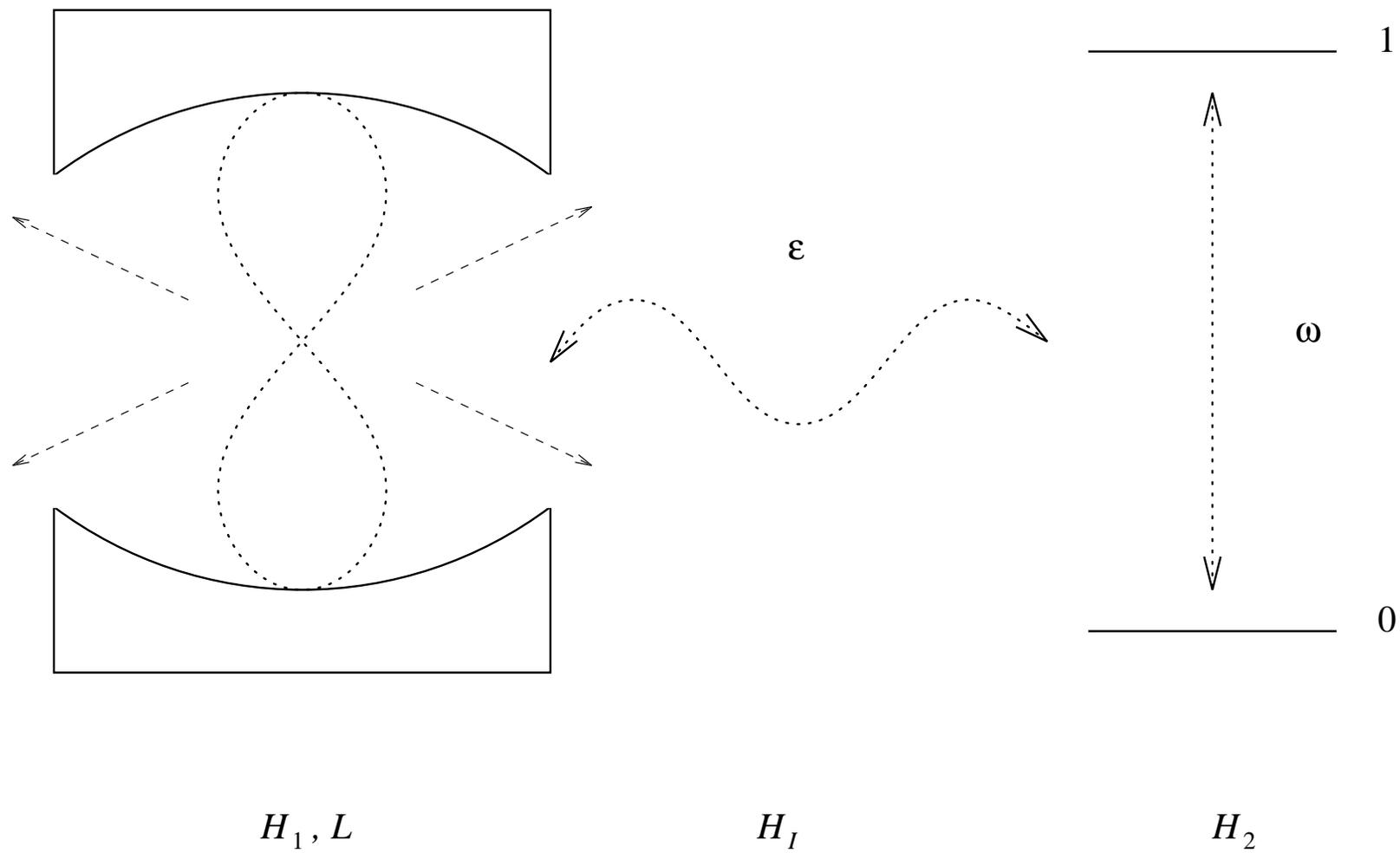
Figure 1.  The interaction between a quantum system with Hamiltonian
$H_1$ interacting with an external field mode
$H_2 = \omega b\d b$ via an interaction potential
$H_I = \epsilon q (b\d + b)$.  The coupling of the system to the environment
is modeled by a single environment operator $L$.  Dotted lines represent
Hamiltonian terms and dashed lines represent coupling to the environment.

\vfil

\end{document}